\documentclass[11pt,twoside]{article}

\usepackage{asp2006}
\usepackage{epsf}
\usepackage{psfig}
\usepackage{graphicx}
\usepackage{lscape}

\begin{document}
\title{The physics and chemistry of circumstellar envelopes of S-stars on the AGB}   %%% Fill in title
\author{S. Ramstedt$^1$, F. L. Sch\"oier$^1$ and H. Olofsson$^{1,2}$}   %%% Fill in author names
\affil{$^1$Stockholm Observatory, AlbaNova University Center, SE-106 91 Stockholm, Sweden \\ $^2$Onsala Space Observatory, SE-439 92 Onsala, Sweden}    %%% Fill in author affiliations

\begin{abstract} %%% Abstract to run on from here.
The S-stars have been suggested to be a brief transitional phase as stars evolve from oxygen-rich M-type stars into carbon stars, through the dredge up of carbon from He-shell burning. As possible transition objects, S-stars might help achieve a deeper understanding of the chemical evolution as a star ascends the AGB, as well as shed more light on the mass-loss mechanism. We have initiated a large survey of 40 S-stars to observe line emission in common molecules such as CO, SiO, HCN, CS and SiS. 
%A good circumstellar model is basic to abundance estimates. 
Detailed radiative transfer modelling of multi-transition CO radio line observations towards a sample of 40 S-stars shows that the mass-loss rate distribution of S-stars is consistent with those found for M-type AGB stars and carbon stars. 
%Possibly there is a trend that the S-stars on average have lower expansion velocities. 
Initial results from modelling of the circumstellar SiO emission are also presented.

\end{abstract}

\section{Introduction}
The molecular setup and grain types in the circumstellar envelopes (CSEs) of AGB stars are to a large extent determined by the C/O-ratio in the photosphere of the central star. There exists three chemical types: the carbon stars, with C/O$>$1, the M-stars, with C/O$<$1, and the S-stars, with C/O$\approx$1. The S-stars are believed to be a short transitional phase, as dredge-up of carbon from He-shell burning, changes the spectral type from M-stars to carbon stars.
The AGB stars contribute to the chemical evolution of galaxies through their extensive mass loss and an understanding of their circumstellar properties is important in the study of galactic chemistry. The mass-loss characteristics and molecular setup of M- and carbon stars are relatively well-studied. However, comparatively little attention has been given to the S-stars. In order to improve this situation we have started a project aimed at obtaining an extensive database of molecular line emission towards a sample of 40 S-stars from common molecules such as CO, SiO, HCN, CS and SiS. 

\section{The physical properties of the CSEs}
The properties of the gas present in the CSEs have been determined from CO data using detailed non-LTE radiative transfer modelling, which self-consistently also calculates the kinetic temperature of the gas \citep{Ramstedt06}. The properties of the dust present in the CSEs were also determined and included in the CO excitation analysis. It was found that the S-stars resembles the mass-loss rate distributions of M-type AGB stars and carbon stars with a median mass-loss rate of  2\,$\times$\,10$^{-7}$\,M$_{\odot}$\,yr$^{-1}$, possibly with a scarcity of objects with high mass-loss rates ($\geq$\,10$^{-5}$\,M$_{\odot}$\,yr$^{-1}$). The expansion velocities of the envelopes are on average similar to the M-type AGB stars but lower than the carbon stars.
\begin{figure}[!ht]
\plotone{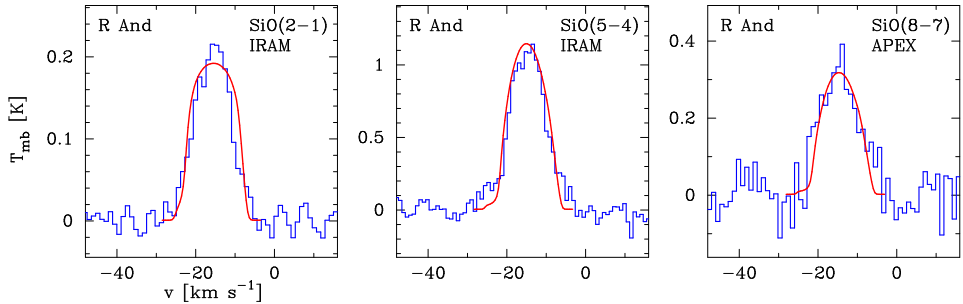}
  \caption{Observed SiO $J$\,=\,2\,$\rightarrow$\,1, \,5\,$\rightarrow$\,4, and \,8\,$\rightarrow$\,7 line emission towards R And (histogram) overlaid with the results from the best-fit model (solid line). The abundance distribution is assumed to follow a single Gaussian with an initial abundance ($f_0$) of 9\,$\times$\,10$^{-6}$ relative to H$_{2}$ and an envelope size ($r_{\mathrm e}$; e-folding radius) of 5\,$\times$\,10$^{15}$\,cm (from Ramstedt et al., in prep.).}
  \end{figure}
\section{The chemistry in the CSEs}
Once the physical properties of the dust and gas present in the CSEs are determined, accurate molecular abundances can be estimated and compared with predictions from chemical modelling, as well as with results for M-stars and carbon stars. In addition to the already published CO data we have detected SiO $J$\,=\,2\,$\rightarrow$\,1 in 20 stars and SiO $J$\,=\,5\,$\rightarrow$\,4 in 18 stars using the IRAM 30\,m telescope. Also, SiO $J$\,=\,8\,$\rightarrow$\,7 has been detected in 17 of the sample stars with APEX. This will be further complimented by observations at JCMT in November 2006. We have also obtained interferometric SiO data for $\Pi^{1}$ Gru and W Aql using ATCA. Moreover, we have gathered data of radio line emission from other molecules, mainly SiS and HCN. 
We have started modelling the SiO radio line emission (see Fig.~1) towards the sample, to determine accurate abundances using a detailed excitation analysis (Ramstedt et al., in prep). Of interest is to see if the S-stars shows the same decline of the average SiO abundance in the CSE with mass-loss rate as is found for M-type AGB-stars \citep{Delgado03b} and carbon stars \citep{Schoeier06a}, indicative of freeze-out on to dust grains. Also, the effect of non-equilibrium chemistry can be further tested.

%As possible transition objects, a study of the S-stars might help achieve a deeper understanding of the chemical evolution as a star ascends the AGB, as well as shed light on the mass-loss mechanism, which is not yet fully understood in detail.
 %
%  \begin{figure}[t]
%  \centering{
%  \includegraphics[width=12cm]{fig1.pdf}
%  \caption{{\bf a)} Mass-loss-rate distributions for the S-star (solid
%    line; 40 stars), M-star (dotted line; 77 stars; \citet{olofetal02}), and carbon star (dashed line; 61 stars; \citet{Schoeier01}) samples.
%    {\bf b)} Envelope gas expansion velocity ($v_e$) distributions for the S-star (solid
%    line), M-star (dotted line), and carbon star (dashed line) samples.
%    {\bf c)} Derived mass-loss rates plotted against $v_e$
%    for the S-star (dots), M-star (squares), and carbon star (triangles) samples.}
%  \label{Mdot}}
%  \end{figure}
%  

%%% MAIN BODY OF TEXT GOES HERE. CONSULT "INSTRUCTIONS FOR AUTHORS USING
%%% LATEX2E MARKUP", SECTIONS 2.3-2.6 FOR HELP WITH EQUATIONS, FIGURES,
%%% AND TABLES.

%\section{}   %%% Top level section head (remove "%" symbol)
%\subsection{}   %%% Second level section head (remove "%" symbol)
%\subsubsection{}   %%% Lowest level section head (remove "%" symbol)
%\section*{}    %%% Unnumbered top level section head (remove "%" symbol)
%\subsection*{}   %%% Unnumbered second level section head (remove "%" symbol)

\acknowledgements %%% Text of acknowledgements runs on after this command.
The authors acknowledge financial support from the Swedish research council.

%%% THE BIBLIOGRAPHY
%%%
%%% CONSULT SECTION 3 OF "INSTRUCTIONS FOR AUTHORS" FOR HOW TO USE NATBIB.
%%% AUTHORS ARE ENCOURAGED TO USE EITHER THE "THEBIBLIOGRAPY" ENVIRONMENT
%%% BY UNCOMMENTING (DELETING THE "%" SYMBOL) THE COMMANDS BELOW, OR BY
%%% USING THE BIBTEX ENVIRONMENT. TO FIND OUT WHICH IS APPLICABLE TO YOUR
%%% CONTRIBUTION, CONSULT THE VOLUME EDITORS FOR YOUR PROCEEDINGS.
%%%

\end{document}